# scientific reports

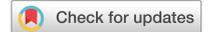

**OPEN**

# Volunteer contributions to Wikipedia increased during COVID-19 mobility restrictions

Thorsten Ruprechter[1]✉, Manoel Horta Ribeiro[2], Tiago Santos[1], Florian Lemmerich[3], Markus Strohmaier[4,5,6], Robert West[2] & Denis Helic[1]

Wikipedia, the largest encyclopedia ever created, is a global initiative driven by volunteer contributions. When the COVID-19 pandemic broke out and mobility restrictions ensued across the globe, it was unclear whether contributions to Wikipedia would decrease in the face of the pandemic, or whether volunteers would withstand the added stress and increase their contributions to accommodate the growing readership uncovered in recent studies. We analyze 223 million edits contributed from 2018 to 2020 across twelve Wikipedia language editions and find that Wikipedia's global volunteer community responded resiliently to the pandemic, substantially increasing both productivity and the number of newcomers who joined the community. For example, contributions to the English Wikipedia increased by over 20% compared to the expectation derived from pre-pandemic data. Our work sheds light on the response of a global volunteer population to the COVID-19 crisis, providing valuable insights into the behavior of critical online communities under stress.

Wikipedia is the world's largest encyclopedia, one of the most prominent volunteer-based information systems in existence[1,2], and one of the most popular destinations on the Web[3]. On an average day in 2019, users from around the world visited Wikipedia about 530 million times and editors voluntarily contributed over 870 thousand edits to one of Wikipedia's language editions (Supplementary Table 1).

Amidst the COVID-19 pandemic and the "infodemic"[4] that ensued, Wikipedia played and continues to play an important role in supplying information about the COVID-19 crisis[5–8]. Notably, although readers accessed medical articles more frequently than in non-pandemic times[9], the increase in readership for all kinds of articles—not only those related to the pandemic—suggests that Wikipedia's role in this time of crisis transcends mere COVID-19-related information seeking[10]. However, page views are but a single aspect of the pandemic's impact on Wikipedia, which ignores the fundamental contribution of editors who perform unpaid volunteer work to maintain and develop content on the website. If the pandemic negatively impacted the productivity and number of editors on Wikipedia, the world's largest online encyclopedia could be in peril[11,12].

We can devise two competing hypotheses on how the COVID-19 crisis may have impacted editors on Wikipedia. First, editing activity on Wikipedia may have declined in response to COVID-19. The negative economic and social ramifications of the pandemic[13–15] that ensued after governments enforced mobility restrictions[16–18] may have adversely affected Wikipedia volunteers. Particularly, the challenges associated with this new reality may have led to an overall decrease in contributions and fewer people joining Wikipedia. For example, potential contributors might have focused their efforts on personal issues and on coping with the crisis rather than volunteering for Wikipedia. Alternatively, volunteer contributions and the number of newly recruited editors may have increased due to mobility restrictions resulting in individuals spending more time at home in front of computer screens[19] or on the Internet[20]. Moreover, as previously observed during locally confined disease outbreaks[21] and extraordinary events[22], the Wikipedia community could have responded to the heightened demand for high-quality information with increased volunteer activity.

Similar to the importance of understanding the response to COVID-19 in other societal or economic contexts[14,15,23,24], we believe that investigating whether volunteer contributions to Wikipedia decreased or increased during this pandemic is critical to assessing the online encyclopedia's ability to continuously provide information to a global audience of readers, even during worldwide disasters.

[1]Graz University of Technology, 8010 Graz, Austria. [2]EPFL, 1015 Lausanne, Switzerland. [3]University of Passau, 94032 Passau, Germany. [4]RWTH Aachen University, 52062 Aachen, Germany. [5]GESIS – Leibniz Institute for the Social Sciences, 50667 Cologne, Germany. [6]Complexity Science Hub Vienna, 1080 Vienna, AT, Austria. ✉email: ruprechter@tugraz.at

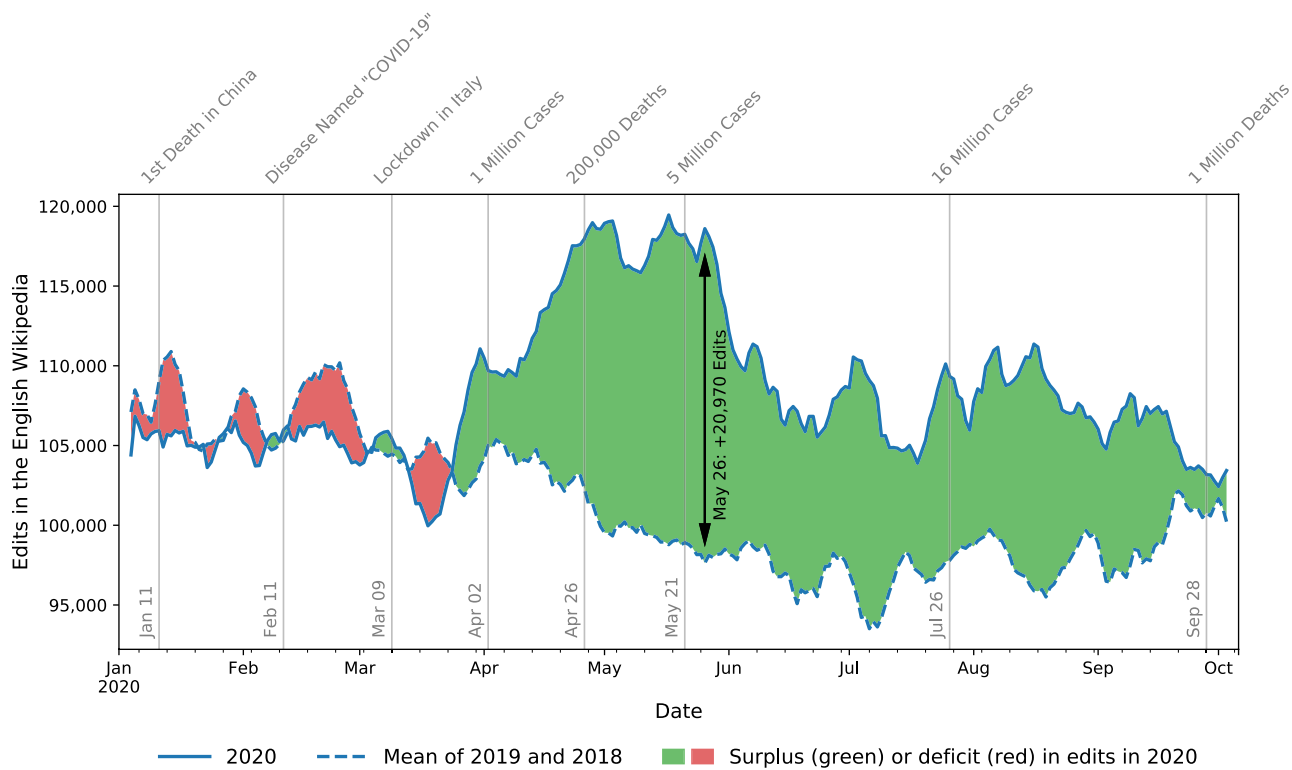

**Figure 1.** Edit volume in the English Wikipedia increased during COVID-19 mobility restrictions. We visualize the rolling 7-day average edit volume in the English Wikipedia from January to October 2020 alongside the daily mean of 2019 and 2018, only considering non-bot edits to Wikipedia articles. Vertical lines mark major developments during the COVID-19 pandemic in 2020 (Via https://wikimediafoundation.org/covid19/data). After the first Western countries (e.g., Italy) enforced mobility restrictions in early March, edit volume stagnated briefly before rising sharply—a trend that prevailed until late May, where the maximum difference in rolling 7-day average edit volume reached 20,970. Although this initial sharp increase in edits later declined, a surplus persisted until October. Until September 31st, editors produced 8.4% (7.3%) more edits in 2020 than in 2019 (2018), an increase of 2.2 million (2 million) edits (Supplementary Table 2). Much of this edit surplus appears to stem from periods of mobility restrictions in the spring of 2020.

After careful quantitative analyses of large-scale edit logs on Wikipedia, we present robust evidence that *volunteer contributions significantly increased during the COVID-19 crisis* across many language editions. During the pandemic, the Wikipedia editor community not only generated many more edits than what we would expect given historical baselines, but also acquired many more newcomers than in recent history, demonstrating the resilience of this online community in the face of adverse conditions.

Figure 1 depicts the increase in volunteer edits in the English Wikipedia during the COVID-19 timeline in 2020 compared to previous years. Whereas no increase in edit volume was apparent in early 2020, the mobility restrictions in Western countries seemed to first slightly dampen edit activity, before triggering a strong upward trend towards the end of March. In the weeks thereafter, a considerable edit surplus developed in comparison to previous years, which lasted until its peak in late May. As the pandemic subsided over the summer, the growth in edit volume also continuously decreased until fall. By October, the relative increase in edit volume, and thus volunteer contribution, from 2019 to 2020 (about 7.9%, or 2.1 million edits) was about double that from 2015 to 2019 (about 4.2%, or 1.5 million edits; see Supplementary Table 2). In summary, a visual representation of edit volume in the English Wikipedia suggests a considerable contribution surplus in 2020.

Beyond the mere descriptive analysis of a single Wikipedia language edition, we systematically analyzed a varied sample of 12 Wikipedia language editions ("Wikipedias"), consisting of four large, medium, and small language editions each ("Methods"), with over 223 million edits spread through 24.6 million articles. In accordance with the descriptive analysis shown in Fig. 1, our quasi-experimental difference-in-differences analysis finds a significant increase in edit volume after COVID-19 mobility restrictions came into effect for many of the Wikipedia editions, and an influx of new editors that is particularly salient for larger Wikipedias. Our study sheds light on the impact of the COVID-19 mobility restrictions on Wikipedia volunteer contributions and provides a reusable framework to measure user activity under stress. More broadly, the evident increase in edit volume and newcomers across most observed Wikipedias is a finding of interest not only to Wikipedia itself but also to researchers and managers of other online collaboration systems, as it provides valuable insight into user behavior during a global crisis.

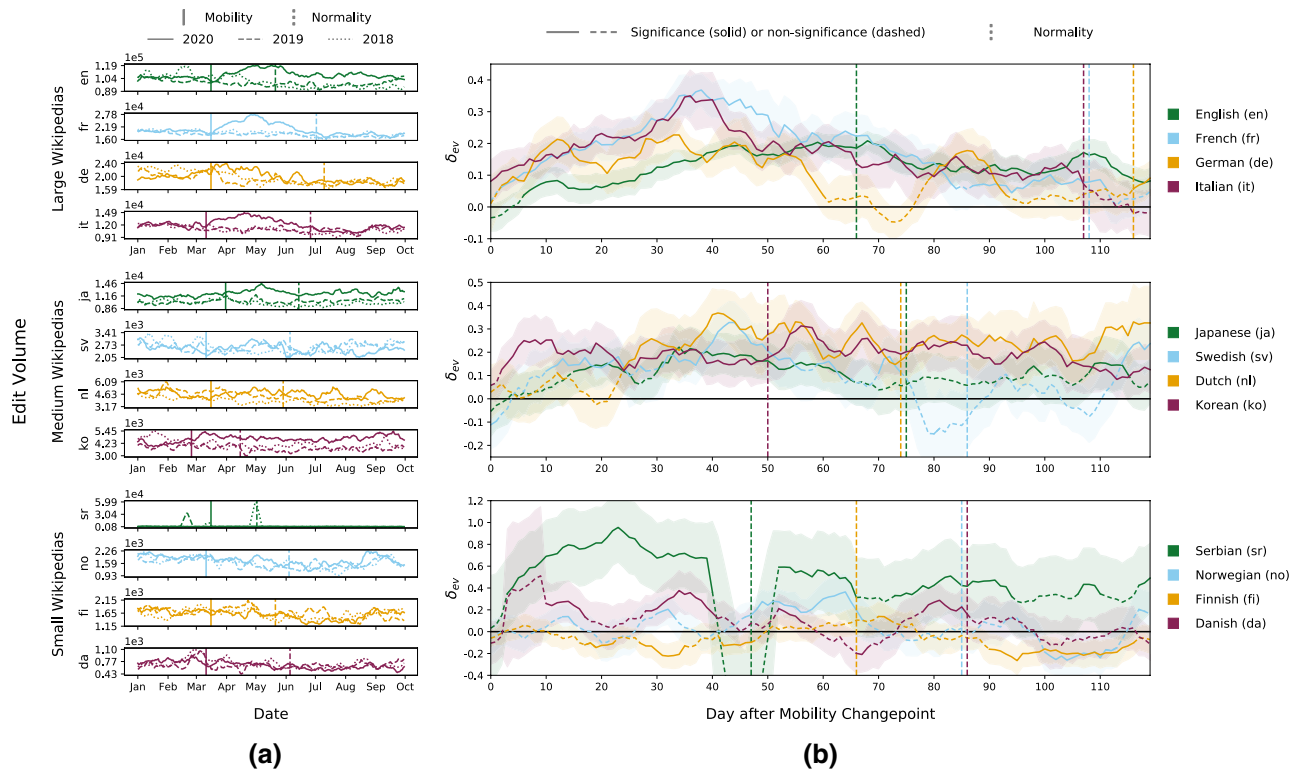

**Figure 2.** Edit volume during COVID-19 mobility restrictions. We show edit volume findings in large (top), medium (middle), and small (bottom) Wikipedias during COVID-19 mobility restrictions, which in the figure we delineate using mobility (when restrictions become effective) and normality (when restrictions are lifted) changepoints. (**a**) We show rolling 7-day average edit volume generated by human editors for 2018, 2019, and 2020 until October. After a slight retraction of editing around mobility changepoints in most Wikipedias, the number of contributions recovers to previous levels within a few days. Editors contribute substantially more in all large and some medium Wikipedias in the weeks after the mobility restrictions in 2020, compared to historical baselines (Serbian Wikipedia experienced several days of high edit activity in May 2018 and February 2019, resulting in the data shown). (**b**) We depict the relative change in edit volume (*ev*) as retrieved from DiD via $\delta_{ev}$ (95% confidence interval as two standard deviations) and plot $\delta_{ev}$ for 120 left-aligned 7-day windows (see "Methods"), with the x-axis describing days after the respective mobility changepoint. We observe that editing in large and medium Wikipedias significantly increases after their mobility changepoint, while most small Wikipedias show neither significant increase nor decrease.

## Results

### Edit volume during COVID-19 mobility restrictions.

We observe an increase in edit volume (the number of edits made by non-bot users) on Wikipedia during the period of COVID-19 mobility restrictions in the spring of 2020, which is particularly evident in large and medium Wikipedias. Figure 2a depicts the rolling 7-day average edit volume for large (top), medium (middle), and small (bottom) Wikipedias in the context of COVID-19 mobility restrictions, which we delineate via automatically detected mobility (i.e., restrictions take effect) and normality (i.e., restrictions are lifted) changepoints (see "Methods"). We also report edit volumes for 2018 and 2019 as a reference for 2020. We observe substantial drops in edit volume around the mobility changepoint for almost all Wikipedias, indicating a shock to the Wikipedia ecosystem. In particular, larger Wikipedias experience a considerable short-lived decrease in edit volume but are able to recover quickly. English, Italian, German, French, Korean, and Japanese even clearly surpass their pre-shock volume levels, leading to an overall edit surplus. On the contrary, some smaller Wikipedias (e.g., Finnish) exhibit a steady decline in edit volume after the mobility changepoint. To better relate edit volume during the COVID-19 pandemic to reference values from previous years and pre-pandemic periods, we employ a difference-in-differences regression (DiD) that controls for the year, period, and language, as well as their interactions. For all Wikipedias, we compute the effective change in edit volume (*ev*) after the mobility changepoint from the three-way interaction of year, period, and language, and denote this effective change as $\delta_{ev}$. We apply the DiD analysis to a sequence of 7-day windows post-changepoint, always retaining the 30-day pre-changepoint period, and plot the time series of logarithmic effects for edit volume according to $\delta_{ev}$ in Fig. 2b. We describe this DiD setup in more detail in "Methods". The DiD analysis validates that all large and most medium Wikipedias significantly increase their edits following the mobility restrictions according to $\delta_{ev}$ (95% confidence interval), while no general statement can be made for small Wikipedias.

For the rolling 7-day average edit volume in large Wikipedias, we identify an upward trend in 2020 immediately after mobility restrictions took place (Fig. 2a, top). In the English, French, and Italian language editions, edit volume steadily increases for nearly 2 months after a dip around the date of the mobility changepoint, before slowly reverting to prior levels. The steady initial increase in edit volume leads to high peaks—approximately 120,000 edits for English, 28,000 for French, 15,000 for Italian, and 24,000 for German, which exhibits a decline back to pre-crisis levels earlier than other large Wikipedias. DiD results confirm the edit volume surplus visible in the time series for large Wikipedias in 2020 (Fig. 2b, top). $\delta_{ev}$ for French, Italian, and English depicts an immediate relative increase in edits after the mobility restrictions take place, leading to over 100 days of significant increases for all three of these Wikipedias, whereas German declines earlier. Approximately 35 days after the mobility restrictions take effect, French ($e^{0.337} = 144\%$, a surplus of 44%), Italian (+42%), and German (+25%) reach their highest significant relative increase for edit volume. The higher short-term increases in French, German, and Italian may be related to more detailed reporting of local issues in these language editions. On the contrary, English shows a longer, sustained upward trend for $\delta_{ev}$, with a maximum relative increase of 23% after 69 days. In conclusion, edit volume significantly increases in large Wikipedias after mobility restrictions come into effect.

Edit volume in most medium and small Wikipedias slightly drops around the respective mobility change-points in 2020. However, virtually all Wikipedias quickly recover from the initial shock, with most maintaining a stable edit volume in the ensuing weeks and some even generating an edit surplus. While Fig. 2a (middle) shows that medium Wikipedias do not homogeneously increase their edit volume, Korean and Japanese surpass their pre-mobility-restriction levels about a month post changepoint, peaking at about 5400 and 14,500 edits, respectively. For small Wikipedias, edit volume only decreases slightly right after the mobility changepoint (Fig. 2a, bottom). Afterward, edit volume recovers to previous baselines within 30 days, before following similar trends and levels as in previous years. DiD analysis and corresponding values for $\delta_{ev}$ reveal that, in fact, medium Wikipedias experience varying periods of significant relative increases in edit volume (Fig. 2b, middle). For example, when compared to pre-pandemic years around the same time period, the Korean and Dutch Wikipedias produce a consistent relative increase (peaking at +40%), whereas Swedish and Japanese exhibit shorter significant periods (+30% and +38% in maximum, resp.). Furthermore, the relative change for small Wikipedias (Fig. 2b, bottom) signals brief periods of substantial relative increases for Danish and Norwegian (peaks of +69% and +43%, resp.). Most notably, Serbian Wikipedia exhibits a considerable increase during the first month after mobility restrictions take place, with volume nearly tripling (logarithmic effect of 1.03). Lastly, we note that out of our twelve investigated Wikipedias only Finnish shows a significant decrease in $\delta_{ev}$ over longer stretches of the observed period. In any case, small and medium Wikipedias are mostly resilient to the initial shock to edit volume triggered by COVID-19, with some even surpassing their pre-pandemic baselines after a few weeks.

**Newcomers during COVID-19 mobility restrictions.** We find that all large and medium Wikipedias acquire considerably more newcomers (the number of registered users who made their first edit) for most of the study period, while the remaining Wikipedias exhibited resilience and do not decrease their levels significantly. We visualize the 7-day rolling averages for newcomer counts during the COVID-19 pandemic for large (top), medium (middle), and small (bottom) Wikipedias in Fig. 3a, while also showing values for previous years as well as mobility and normality changepoints. Newcomer counts plummet around the mobility changepoint, in particular for large Wikipedias, but this attenuation in newcomer recruitment only persists for a brief period. Shortly thereafter, newcomer counts increase considerably in all but a few medium and small Wikipedias (e.g., Swedish or Finnish). Again, we build a DiD model for newcomers ($nc$) to quantify effective changes during the period of COVID-19 mobility restrictions in spring 2020, controlling for year, period, and language. We again perform our DiD analysis for a sequence of 7-day windows after the mobility changepoint (see "Methods") and show the logarithmic effects for newcomers ($\delta_{nc}$) in Fig. 3b. This newcomer DiD analysis confirms that while all large Wikipedias acquire significantly more new editors after mobility restrictions take effect, some medium and small Wikipedias seem to be resilient and exhibit no significant long-term changes (95% CI).

Large Wikipedias appear to recover rapidly from the initial negative effect of mobility restrictions in terms of newcomer counts (Fig. 3a, top). Most notably, Italian Wikipedia registers a newcomer surge until late April, recruiting over 150 newcomers on a rolling 7-day average. English and French show similar patterns of perpetual increases, reaching respective peaks of approximately 2100 and 330 new editors. Although German exhibits nearly 200 newcomers shortly after the mobility changepoint, the surplus in 2020 seems not as considerable as for other large Wikipedias. We further note that newcomer counts for large Wikipedias start to steadily decline in May. However, this seasonal trend also appears to be prevalent in previous years. Our DiD analysis, which captures the change in newcomers via $\delta_{nc}$, for the most part confirms these findings (Fig. 3b, top). During the first 2–3 weeks past the mobility changepoint, large Wikipedias steadily recover from the COVID-19 shock without significant overall gains according to $\delta_{nc}$. However, right after this recovery phase significant peaks arise for English ($e^{0.283} = 130\%$ of previous levels), French (138%), and German (139%). For the Italian Wikipedia, which belongs to a region with particularly strict mobility restrictions, we confirm an even stronger newcomer surge, leading to a 80% relative increase. Furthermore, English generates a rather stable, significant long-term growth in newcomers that is possibly owed to editors from all over the world joining this language edition during mobility restrictions in their regions, as the English Wikipedia serves as a global repository of knowledge. Ultimately, positive effects prevail for large Wikipedias and solidify a newcomer surplus after the mobility restrictions come into force.

Similar to large Wikipedias, most medium and small Wikipedias experience a decline in newcomers right around their mobility changepoints before then increasing their counts to previous baselines (Fig. 3a, middle and bottom). Some of these Wikipedias (e.g., Norwegian, Finnish, Danish, Swedish) recover to previous levels within the first month and exhibit no long-term effects afterwards. However, others recruit a surplus of newcomers during this crisis. Japanese, Dutch, Korean, and Serbian show short-term newcomer influxes about 1–2 months after the initial mobility restrictions take effect, with maximum respective values of approximately 170, 60, 50, and 30 daily newcomers. We also observe these effects in $\delta_{nc}$ as captured by DiD (Fig. 3b, middle and bottom), which confirms brief relative increases for Japanese (+30%), Dutch (+47%), Korean (+28%), and Serbian (+179%). Finally, the newcomer DiD analysis corroborates that some medium and most small Wikipedias do not significantly deviate from baselines prior to the mobility restrictions over much of the observed time span.

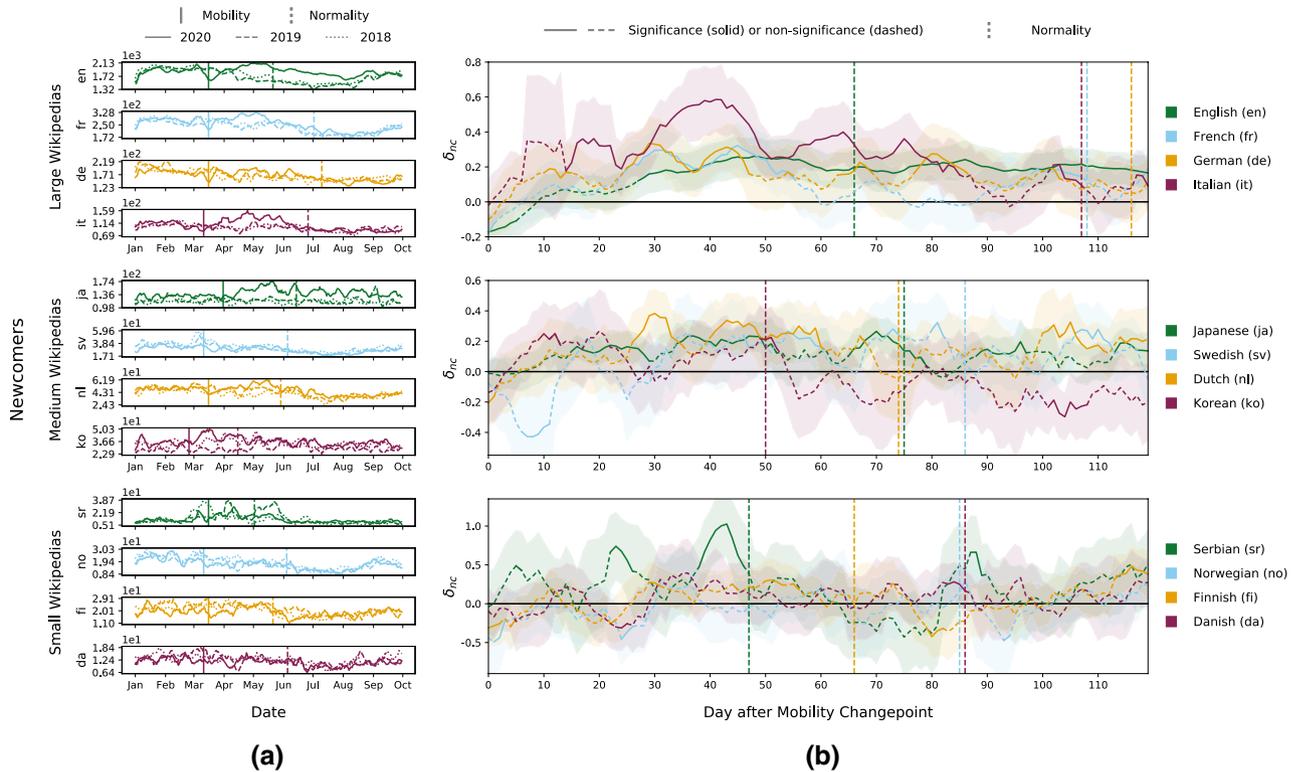

**Figure 3.** Newcomers during COVID-19 mobility restrictions. We visualize newcomer results in large (top), medium (middle), and small (bottom) Wikipedias during COVID-19 mobility restrictions, which we delineate via mobility (when restrictions become effective) and normality (when restrictions are lifted) changepoints. (**a**) We depict rolling 7-day average newcomer counts until October of 2018, 2019, and 2020. For many Wikipedias, newcomer acquisition strongly declines right around their mobility changepoint, but then quickly rises to or even exceeds pre-pandemic baselines. (**b**) We investigate the relative change in newcomers (*nc*) via $\delta_{nc}$ as computed from DiD analysis (95% confidence intervals as two standard deviations) and plot $\delta_{nc}$ for 120 left-aligned 7-day windows (see "Methods"), starting with the respective mobility changepoint. In large Wikipedias, considerably more newcomers join in the weeks after mobility restrictions come into effect, relative to before the changepoint and previous years. Results for medium and small Wikipedias are non-conclusive, with some showing increases in the number of newly acquired editors and others not significantly changing their values.

## Discussion

As the COVID-19 pandemic erupted on a global scale, it was unclear how this incisive event would affect Wikipedia's volunteer community. Over the course of the last few years, both human editing[25] and newcomer recruitment[26] on Wikipedia have stagnated or even decreased (Supplementary Table 2). Accordingly, the pandemic could have accelerated the decline of the online encyclopedia as the hardships of this global crisis may even further decrease volunteer activity. However, our study, in which we analyze 223 million edits from 12 Wikipedia language editions, reveals that the COVID-19 pandemic and its accompanying mobility restrictions have substantially boosted volunteer activity on Wikipedia. By performing a difference-in-differences analysis, we show that edit volume as well as the influx of newcomers has generally *increased* after COVID-19 mobility restrictions went into effect. In what follows, we discuss the implications and limitations of this finding.

**Mechanisms behind contribution growth.** We observe significant increases in edit volume and newcomers during the COVID-19 pandemic across multiple Wikipedias, making it their most active period in at least the last 3 years. While our quantitative study sheds light on the extent of contribution growth, there are several possible mechanisms behind this effect, which may or may not impact the collaborative structure of editor communities.

Firstly, Wikipedia received significantly more page views during the COVID-19 crisis[10]. The increase in edits and newcomers may partially be due to the prior increase in Wikipedia readership, as a certain proportion of readers turns into contributors because of various motivational factors[27,28]. In addition, we theorize that increased screen time and Internet exposure[19,20] during the mobility restrictions lead to Wikipedia readers spending more time editing, possibly increasing the reader-to-editor turnover rate. Tracing the transformation of readers into editors during this pandemic in more detail is a promising avenue for future work.

Secondly, the increase in contributions may be due to the rapidly changing information and new knowledge that the COVID-19 pandemic generates about the world. Past literature has suggested that Wikipedia growth is constrained by the amount of knowledge available, as editors have already contributed most of the easily obtainable and verifiable information[25]. The fact that volunteers have been "running out of easy topics" to contribute to has made it difficult for non-specialists to provide new content with little effort[11]. As the COVID-19 pandemic dramatically changes the status quo of our world today, it is generating new knowledge about many fields and thus may provide fresh opportunities for both novel and veteran editors to contribute to Wikipedia.

Moreover, the observed edit surplus may have been caused by the high-intensity activity of a core group of editors rather than the broader editor population. We therefore investigate the number of editors active on any given day according to their activity level: 1–4, 5–24, 25–99, or more than 99 daily edits ("Methods"). The DiD analysis for editor counts depicts increases across all activity levels after mobility changepoints for all large and most medium Wikipedias, while small Wikipedias show non-conclusive effects (Supplementary Figs. 1, 2, 3 and 4). This corresponds to an overall increase in active editors during the pandemic (Supplementary Fig. 5), indicating that the general editor population intensified its contribution during the COVID-19 pandemic.

On top of that, we detected a contribution disparity with respect to Wikipedia size, meaning that the smaller Wikipedias we studied did not benefit to the same degree as larger or medium Wikipedias. The observed discrepancy in edit and newcomer increases for large, medium, and small Wikipedias may stem from a difference in community size and structure, or these Wikipedias' specific rules[11,29–32]. This discrepancy could be further reinforced by the culture and perception of what constitutes good knowledge representation in the different Wikipedia language editions[33]. Moreover, the amount of content for certain topical categories diverges due to cultural contextualization in different language editions[34]. Specifically, a strong (hypothetical) affinity for topics not directly related to the pandemic (e.g., *Sports*) in medium or smaller Wikipedias might change the effect of this crisis on their edit volume, in comparison to larger Wikipedias. As an example, in case such Wikipedia language editions focused more on updating sports articles, edit volume would decrease more during the pandemic. The magnitude of such an effect may further depend on a region's more (e.g., Italy) or less strict (e.g., Sweden) mobility restrictions. Future research may explore language-specific collaboration mechanisms in more detail, for example by attempting to topically analyze Wikipedia contributions during the pandemic.

Lastly, in this study we investigate contribution growth in terms of the number of edits and newcomers, rather than alternative measures such as the total sum of contributed or surviving content. To also touch on this aspect of Wikipedia contributions, we coarsely analyze the contributed information in bytes during COVID-19 mobility restrictions (Supplementary Figure 9). While the surplus in contributed content largely does not match the surplus found for edits and newcomers, we observe an increase for several languages after the imposed mobility changepoints. Furthermore, we find no significant decreases for all Wikipedia editions. It might prove fruitful to extend this quantitative analysis by examining the actual dynamics and structure of the contributed information.

**Resilience of Wikipedia communities.** Although we did not find the same surplus in contributions across large, medium, and small Wikipedia language editions, volunteer communities in all studied Wikipedias demonstrated resilience by quickly recovering from the initial negative impact of the pandemic on their contributions. While slow response to negative events or other shocks causes severe problems in social-ecological systems[35,36], resilient systems are adaptable and manage to withstand such shocks, even bearing the capacity to cross previous performance thresholds[37]—a behavior observed in this study. The strongest resilience and subsequent crossing of earlier thresholds in large Wikipedias during the pandemic may be partially explained by the difference in community size[38]. For example, in larger communities it may not be as problematic that leaders are limited due to the pandemic, as a greater number of other veteran members can take over their work. This conjecture borrows from critical mass theory, in the sense that a critical mass of core members is the fundamental source of content[27]. Future research might investigate the aspect of Wikipedia resilience during the pandemic in more detail, for example by considering threat rigidity[38] or building a model[39] that considers COVID-19 as an attack on the community structure.

**Revert rate during COVID-19 mobility restrictions.** The observed simultaneous increase in newcomers and edits may indicate that the edit surplus was partially caused by first-time editors. Although past research has suggested that new or one-time editors often produce high-quality contributions[40], veteran editors or bots would frequently completely undo (i.e., identity revert) these newcomer revisions, which represents a common behavioral pattern on Wikipedia[11,26,41] that in turn generates further revisions. To investigate whether an increase in reverts occurred, we performed a cursory analysis of the revert rate, which is defined as the ratio of reverted edits to edit volume (see "Methods" and Supplementary Information). Supplementary Figure 6a visualizes the rolling 7-day average revert rate, while Supplementary Figure 6b plots the relative change in revert rate as captured by a DiD analysis ("Methods"). It is noteworthy that we only detect a significant increase of the revert rate in one language (Korean). By contrast, several Wikipedias exhibit significantly decreased revert rates shortly after the mobility restrictions come into force. For example, the large Italian, French, and German Wikipedias all show reduced revert rates by about one quarter. This suggests that less valuable revisions, possibly made by newcomers, and their immediate reversal do not cause the reported increase in edit volume. Furthermore, potential misbehavior or conflict on Wikipedia, such as vandalism or edit wars, is prominently characterized by large numbers of identity reverts, as they undo these unwanted contributions[42–46]. Therefore, reduced revert rates may indicate that editors refrain more from confrontational behavior and thus demonstrate higher levels of solidarity during the pandemic, which is a common phenomena within collectives during crises[47]. However, a decline in revert rate could also imply that bots and administrators may be unable to keep up with the influx of edits, leaving low quality or malicious edits undetected and thus diminishing quality in the long term. Moreover, differences in revert rates across Wikipedia editions could also be explained by divergent page protection and blocking policies, where either certain groups of editors or even all public editors are prevented from editing articles that are subject to increased attention. We see the detection and analysis of behavioral patterns and collaborative structure of online communities as a promising path for future research. Additionally, it may be valuable to further study the treatment and retention of newcomers[26,30,48] during and after the pandemic once more longitudinal data is available.

**Contribution to COVID-19 articles.** One might speculate that the increase in edit volume is mostly due to edits in articles that are strongly related to COVID-19. However, many of those articles were protected from public editing early in the pandemic to prevent spread of misinformation[49], and we find that only a negligibly small fraction of edits (at most 1% for most Wikipedias) goes towards articles with a primary focus on COVID-19 (see "Methods") between January 1st and September 31st 2020 (Supplementary Table 4; Supplementary Fig. 7). A clear outlier in that regard is German, where 2.5% of edits performed in 2020 by the end of September concern themselves with such articles. This may indicate higher coverage of local COVID-19 outbreaks in German than in other languages. We consequently repeat our DiD analysis for edit volume, this time excluding edits to articles strongly related to COVID-19 (Supplementary Fig. 8). The results support the previous findings and confirm that the reported edit volume increase is not due to COVID-19 articles. In this way, our work extends previous studies, which focused on a smaller subset of pandemic-related articles[7,49].

**Other limitations.** Even though our work covers a large portion of Wikipedia's content and editor population, it comes with several limitations. First, we do not consider a variety of different Wikipedias associated with languages widely spoken in the Global South, including Spanish, Portuguese, Arabic, Hindi, or any African Wikipedias (see "Methods" for how we chose language editions). Future work analyzing these Wikipedias could improve our understanding of the impact of the pandemic on volunteer contribution in other parts of the world. Second, content on Wikipedia is predominantly edited by white males between the ages of 17 and 40[50,51]. It may be that the COVID-19 crisis has disparately impacted contributors of less represented demographics, as certain racial or socioeconomic groups are particularly disadvantaged by the pandemic[23,24,52]. In addition, bots have an important role in the creation and management of Wikipedia content[41,53]. We excluded bots from our analysis as we specifically focused on edits performed by human volunteers. Nevertheless, other studies may choose to consider bot activities as valid contributions to Wikipedia.

In conclusion, our study provides evidence for a substantial surplus of volunteer contributions to multiple Wikipedia language editions during COVID-19 mobility restrictions, which shines light on the resilience of the Wikipedia community under times of stress. The methodological framework used in this work can easily be adapted for similar domains. We believe that our work provides valuable insights into contributor behavior on online platforms during the COVID-19 pandemic and illustrates a plethora of possibilities for future work.

## Methods
**Data procurement and preprocessing.** We utilize the openly available MediaWiki history dumps dataset[54] to analyze a varied sample of 12 Wikipedia language editions ("Wikipedias").

*Wikipedia language editions.* We investigate 12 Wikipedias (Supplementary Table 3), consisting of languages primarily spoken in European countries that were exposed to the outbreak of COVID-19 in the spring of 2020, as well as two Asian Wikipedias. Our choice of language editions takes into consideration: (1) the size of the Wikipedia edition, (2) whether the language is spoken in relatively few countries, and (3) the mobility restrictions imposed in these countries—three criteria that are often very difficult to simultaneously satisfy. Overall, we aim to capture relevant Wikipedias that represent different attitudes towards the crisis, preferably from languages easily attributable to a single country or region. Accordingly, our sample contains regions with strict (e.g., Italian, Serbian, or French) and less stringent mobility restrictions (e.g., Japanese, Korean, or Swedish). Although it can not be attributed to a single country and about half of the contributions come from editors who are not based in English speaking regions[55], we include the English Wikipedia because it is the largest language edition and is considered a global project. We employ the number of edits in 2019 as a metric to categorize the 12 Wikipedias we studied as either *large* (English, French, German, Italian, with more than 5 million edits), *medium* (Swedish, Korean, Japanese, Dutch, with 1.5 million to 5 million edits), or *small* (Serbian, Norwegian, Danish, Finnish, with less than 1.5 million edits).

Our work only covers an arbitrary part of the Wikipedia ecosystem, as we have argued above. To allow extension of our analysis to other Wikipedia language editions or Wikimedia projects (e.g., Wikidata, Wikimedia Commons, or Wiktionary), we make our code publicly available (see Code availability). As an example, we have provided a demonstration in our GitHub repository for adapting our code pipeline to explore Wikipedia editions covering large parts of Eastern Europe (Polish, Czech, Ukrainian, and Russian Wikipedia).

*MediaWiki history dataset dumps.* We retrieve the monthly updated MediaWiki history dataset dumps[54] provided by the Wikimedia Foundation (WMF) and perform additional preprocessing before computing as well as plotting our results. The denormalized MediaWiki history dumps are generated from the full history logs stored in the WMF's MediaWiki databases. During their generation, WMF's automatic scripts reconstruct and enrich user and page history with additional data, and also automatically validate the dumps to prevent errors. After WMF's preprocessing, the dataset contains fields with precomputed standard metrics, such as revert information, bot users, number of user contributions, or time since a user's last revision. Overall, each entry in the dump consists of 70 fields with event information. Fields are grouped into entities, bearing information about either `revision`, `page`, or `user`.

*Preprocessing.* In the MediaWiki history dataset, we only consider edits to articles by excluding all pages not in the Wikipedia article namespace ("ns0"), thus removing revisions to talk pages or other content. Furthermore, we utilize corresponding dataset fields to distinguish human editors (anonymous or registered) from bots and mark certain revisions as reverts. Moreover, we convert MediaWiki history timestamps from Coordinated Universal Time (UTC) to the timezone of the local Wikipedia language edition. For Wikipedias in which languages can not be attributed to a single timezone (e.g., French), we choose the timezone with the highest volunteer population for the given Wikipedia. We do not apply timestamp conversion for the English Wikipedia. Lastly, we detect articles which are strongly related to COVID-19 via an algorithm by Diego Sáez-Trumper[56], which recognizes COVID-19 articles based on their Wikidata[57] links to the main COVID-19 pages.

**Metrics.** To make sense of which exact data fields in the MediaWiki history dumps we utilize to compute our metrics, please refer to the code repository (see Code availability).

*Edit volume.* We define edit volume as the number of daily revisions to pages in the article namespace ("ns0") by non-bot users (anonymous or registered).

*Newcomers.* For each Wikipedia language edition and day, we specify the amount of newcomers as the number of registered editors which perform their first article edit in that language edition on the given day. Through recognizing new editors by their first edit, we measure the exact day they become a contributor in a language edition. Note that the number of daily registered users is generally much higher than the number of newcomers as computed in this work. However, as our study aims to quantify volunteer contribution, we choose to identify newcomers by their first actual contribution in a given Wikipedia.

*Revert rate.* Editors and bots revert article revisions to undo changes which they deem unwarranted. Frequently, these reverts correct revisions which arise from conflicts, edit wars, or vandalism[42]. Additionally, literature shows that revisions by newcomers are more likely to be reverted than those of veteran editors[26]. For this research, we only consider reverts to articles that undo all changes and subsequently create a new revision which exactly matches a previous article version (i.e., identity reverts). We calculate the daily revert rate by dividing the number of identity reverts (by humans or bots) by the number of non-bot edits on this given day. Correspondingly, revert rate relates the amount of reverts to the amount of human contribution.

*Daily editors by activity level.* We measure daily active editors in a Wikipedia by counting the number of registered, non-bot users which perform revisions in the article namespace. In addition, to detect effects across the editor population, we collect data for multiple activity levels, keeping count of how many editors perform 1–4, 5–24, 25–99, or more than 99 daily edits. In contrast to other metrics, we do not compute the number of daily editors from the Wikimedia history dumps, but retrieve it via the Wikimedia REST API instead[58].

*Contributed information in bytes.* As a parsimonious representation of the amount of information editors contribute to Wikipedia, we measure the daily changes in bytes to all articles in each language edition by summing up the difference in bytes between each revision and its parent revision. We include bot edits to account for the automated correction of vandalistic content additions and removals, which would otherwise distort the results. Although this metric does not provide information about the actual value (or longevity) of the contributions, it does provide an overview of the editors' workload on any given day.

**Changepoint detection.** We adopt the approach by Horta Ribeiro et al.[10] to detect *mobility* and *normality* changepoints via Google and Apple mobility reports, as detailed in Supplementary Figure 10. The mobility reports capture population-wide movement patterns based on cellphone location signals and specify, on a daily basis, the percentage of time spent in variety of locations, for example residential areas, workplaces, or retail (https://www.apple.com/covid19/mobility; https://www.google.com/covid19/mobility). Government-mandated lockdowns and self-motivated social distancing measures manifest themselves as sharp changes in these mobility time series. To detect changepoints in mobility, the approach consists of a simple binary segmentation algorithm[59]. For Wikipedias of languages widely spoken across many countries (e.g. English, German, etc.), we determine a changepoint by aggregating mobility reports for the countries in which the language is official with weights proportional to the population of each of these countries. Notice that the link between Wikipedia and language editions is merely approximate—in particular for English, which is accessed from all over the world. We use the changepoints at which mobility drops as heuristics for dates when people started spending substantially more time in their homes and term them *mobility changepoints*. To detect *normality changepoints*, we compute the point in time for which the future average mobility remains within a 10% band around baseline levels before the initial mobility changepoint (defined as pre-pandemic mobility levels by Google and Apple). For languages spoken across multiple countries, we maintain the same aggregation scheme as before. Compared to choosing specific dates, this changepoint detection approach leads to more comparable treatments across different regions. Supplementary Table 3 summarizes the detected changepoints, which we also make available in our code repository.

While we solely utilize the *normality changepoints* as visual guidance in the figures, our employed causal inference strategy described below strongly relies on the *mobility changepoints*. Therefore, we perform a sensitivity analysis with varying mobility changepoints for all main results. We vary the mobility changepoints by $\pm 7$ days and also consider a distinct method of aggregation for languages spoken in multiple countries (Supplementary Fig. 10). As the latter method shows a maximum difference of 5 days across all languages, we only report results using the 7-day difference.

**Difference-in-differences setup.** To compare values of metrics during the COVID-19 pandemic with reference values from previous years and pre-pandemic periods, we employ a difference-in-differences regression (DiD). DiD analysis allows us to quantify changes in these metrics in multiple Wikipedia language editions around times of region-specific mobility changepoints in early spring, while controlling for (long-term) temporal trends.

Our basic DiD equation models a dependent variable's value ($V$) as a function of the independent variables year ($Y$), period ($P$), and Wikipedia language ($L$), as well as their interactions. Year is a binary variable which differentiates between pre-pandemic (2018 and 2019) and pandemic years (2020), whereas period encodes the treatment period via a binary variable, in our case represented by the pre- and post-phases of the region-specific mobility changepoints. Lastly, we model our 12 Wikipedia language versions with a categorical variable to control for language-specific effects. To account for outliers and normalize regression results across various-sized Wikipedias, we use logarithmic scales for $V$. Literature often refers to our setup, which uses three independent variables, as "triple-difference" or "difference-in-difference-in-difference" estimators[60,61]. Mathematically, our DiD setup is:

$$V = \beta_0 + \boldsymbol{\beta_1}^\top L + \beta_2 Y + \beta_3 P + \boldsymbol{\beta_4}^\top (YL) + \boldsymbol{\beta_5}^\top (PL) + \beta_6 (YP) + \boldsymbol{\beta_7}^\top (YPL) + \varepsilon \qquad (1)$$

We depict the 12 Wikipedia language versions as a vector of 11 binary indicators ($L$). Scalar coefficients ($\beta_0$, $\beta_2$, $\beta_3$, $\beta_6$) describe effects for the reference language (i.e., baseline). Coefficient vectors ($\boldsymbol{\beta_1}$, $\boldsymbol{\beta_4}$, $\boldsymbol{\beta_5}$, $\boldsymbol{\beta_7}$, printed in bold) collect language-specific effects of non-baseline Wikipedias. Lastly, $\varepsilon$ is the normally distributed residual. Given this mathematical formulation, the coefficient $\boldsymbol{\beta_7}$ captures the change in $V$ post mobility changepoint relative to the baseline Wikipedia, after accounting for differences stemming from year or period alone ($\boldsymbol{\beta_4}$ and $\boldsymbol{\beta_5}$, resp.). We therefore compute the effect of interest for all Wikipedias via summation of $\beta_6$ and $\boldsymbol{\beta_7}$. For each Wikipedia, we term this effective change in $V$ as $\delta_m$, where $m$ stands for the metric representing the dependent variable.

*Interpretation of DiD coefficients.* We now elaborate in more detail on how to interpret the coefficients of our DiD model. We model the categorical language variable via vector $L$ containing 11 binary indicator variables for the 12 Wikipedias. As is customary, the regression utilizes a "reference Wikipedia" baseline. In our setup, we arbitrarily choose Danish as the baseline. Consequently, $\boldsymbol{\beta_1}$ describes the respective difference between the baseline Wikipedia and the 11 non-baseline Wikipedias using indicator variables. Thus, adding $\beta_0$ and $\boldsymbol{\beta_1}$ yields the intercept of each language's sub-model.

The binary year variable ($Y$) indicates whether a data point lies in 2020 ($= 1$) or in the previous two pre-pandemic years ($= 0$), regardless of period. As Danish represents the arbitrary baseline, the corresponding coefficient $\beta_2$ is a scalar which describes the overall change between the pre-pandemic years (2018 and 2019) and 2020 for Danish. For non-baseline Wikipedias, the interaction $YL$ models the language-specific effects for the change in years relative to the baseline Wikipedia and is quantified by the corresponding coefficient vector $\boldsymbol{\beta_4}$. Therefore, the summation of $\beta_2$ and $\boldsymbol{\beta_4}$ is equal to the effective overall difference of 2020 to the previous 2 years for all Wikipedias.

We model seasonal differences between pre- and post-changepoint windows via the binary period indicator ($P$). The corresponding scalar coefficient ($\beta_3$) measures the difference between before and after the mobility changepoint over all years for the baseline. Consequently, $PL$ and coefficient vector $\boldsymbol{\beta_5}$ describe the period effect for non-baseline Wikipedias in relation to the baseline. Calculating the sum of $\beta_3$ and $\boldsymbol{\beta_5}$ then gives the total pre- and post-changepoint effects.

Lastly, the interaction between year and period ($YP$) enables our model to capture the change in $V$ for the baseline Wikipedia via $\beta_6$, after accounting for change in $Y$ (via $\beta_2$) and $P$ (via $\beta_3$) alone. To measure this effective change for all Wikipedias, we employ the coefficient vector $\boldsymbol{\beta_7}$ of the three-way interaction $YPL$. While $\beta_6$ describes the baseline's effect, $\boldsymbol{\beta_7}$ contains the aforementioned change relative to the baseline Wikipedia. Therefore, the sum of $\beta_6$ and $\boldsymbol{\beta_7}$ captures the effective change in $V$ for all Wikipedias. For a single Wikipedia and metric $m$, we name this effect of interest $\delta_m$. Correspondingly, $\delta_m$ describes language-specific post-changepoint effects in 2020, as it excludes differences that are due to year or period alone.

**Quantifying changes in volunteer contribution.** Wikipedia is a dynamic ecosystem, in which edit behavior and the amount of volunteer contribution can change rapidly—especially in times of turmoil. To track these changes and detect short-, medium-, and long-term effects of mobility restrictions on volunteer contributions, we fit our statistical model on different data-points obtained from the same longitudinal dataset. This methodology, pioneered by Gelman and Huang[62], allows us to observe trends rather than mere point estimates.

We compute our DiD analysis for a sequence of post-changepoint windows, always retaining the Wikipedias' pre-changepoint periods. For each language version, we choose a fixed 30-day period before the respective mobility changepoint as the pre-changepoint baseline. As post-changepoint analysis intervals, we then extract a sequence of 120 overlapping left-aligned 7-day windows starting with the changepoints. Mathematically, we set the treatment period to days $\{n, n+1, \ldots, n+6\}, \forall n \in \{0, 1, \ldots, 119\}$. For each post-changepoint window $n$, we perform a separate DiD analysis across all languages using the retained baseline periods. By doing so, each DiD analysis compares the week starting at day $n$ after the language-specific changepoint to the baseline periods. In this default setup, each of the 12 Wikipedias is represented by 37 data points for every year in the DiD regression (2018, 2019, and 2020), yielding a total of 1332 data points ($= (30$ pre-changepoint days $+ 7$ post-changepoint days$) \times 3$ years $\times 12$ Wikipedias) for each of the 120 experiments. For each Wikipedia, we conservatively detect outliers via the Median Absolute Deviation (MAD) approach[63] with a threshold of $5 * \text{MAD}$ from the monthly median and replace such outliers by the monthly median. We then build a time series of the 120 DiD results using $\delta_m$ and approximate the 95% two-sided confidence intervals (CI) as two standard errors. As robustness checks, we compute variations of our DiD experiments with wider window size (14 days) and slightly varied mobility changepoint dates ($\pm 7$ days) as described in Supplementary Information (Supplementary Figs. 11, 12, 13, 14, 15, 16, 17, 18, 19). These results corroborate the findings reported under "Results".

*Identifying assumptions of DiD estimators.* Triple-difference and difference-and-difference models in general underlie strong identifying assumptions. In particular, our methodology is based on the assumption that the difference between treatment and control group would stay constant without an intervention[64]. Additionally, the condition of parallel trend in slopes prior to the treatment must be fulfilled[65]. For our model, we therefore assume that for each Wikipedia language version, the difference between treated (i.e., 2020) and control years (i.e., 2018 and 2019) would remain constant in the absence of an intervention (i.e., mobility changepoint). Furthermore, our model trivially fulfills the parallel trend condition: Given that our model only considers a single period before ($P = 0$) and after ($P = 1$) the intervention, testing for parallel trends prior to the intervention[64] results in failing to reject the null hypothesis (that there are no violations of parallel trends). In other words, as these single pre-intervention data points do not produce trends prior to the changepoints, they do not, by definition, deviate significantly from parallelism. Finally, as explained above, we fit several DiD estimators to derive longitudinal intervention effects, which provides additional robustness with respect to the assumptions of such models.

## Data availability

The openly accessible MediaWiki history dataset dumps are available at https://dumps.wikimedia.org/other/mediawiki_history/readme.html[54]. We further provide preprocessed data and results relevant to the manuscript in the code repository at https://github.com/ruptho/wiki-volunteers-covid.

## Code availability

The code repository for this paper can be found at https://github.com/ruptho/wiki-volunteers-covid.

## References

1. Giles, J. *Internet Encyclopaedias Go Head to Head* (Nature Publishing Group, 2005).
2. Lemmerich, F., Sáez-Trumper, D., West, R. & Zia, L. Why the world reads Wikipedia: Beyond english speakers. In *Proceedings of the Twelfth ACM International Conference on Web Search and Data Mining* (2019), pp. 618–626.
3. Alexa. The top 500 sites on the Web. https://www.alexa.com/topsites. Accessed 07 Jan 2021 (2021).
4. Gallotti, R., Valle, F., Castaldo, N., Sacco, P., & De Domenico, M. Assessing the risks of infodemics in response to COVID-19 epidemics. arXiv:2004.03997 (arXiv preprint) (2020).
5. Colavizza, G. COVID-19 research in Wikipedia. *bioRxiv* (2020).
6. World Health Organization. The world health organization and wikimedia foundation expand access to trusted information about COVID-19 on Wikipedia (2020).
7. Gozzi, N., *et al.* Collective response to the media coverage of COVID-19 pandemic on Reddit and Wikipedia. arXiv:2006.06446 (arXiv preprint) (2020).
8. Wikimedia Foundation. Responding to COVID-19: How we can help in this time of uncertainty (2020).
9. Chrzanowski, J., Sołek, J., Fendler, W. & Jemielniak, D. Assessing public interest based on Wikipedia's most visited medical articles during the SARS-CoV-2 outbreak: Search trends analysis. *J. Med. Internet Res.* **23**(4), e26331 (2021).
10. Horta Ribeiro, M. *et al.* Sudden attention shifts on wikipedia during the covid-19 crisis. *Proc. Int. AAAI Conf. Web Soc. Media* **15**(1), 208–219 (2021).
11. Halfaker, A., Geiger, R. S., Morgan, J. T. & Riedl, J. The rise and decline of an open collaboration system: How Wikipedia's reaction to popularity is causing its decline. *Am. Behav. Sci.* **57**(5), 664–688 (2013).
12. Ransbotham, S. & Kane, G. Membership turnover and collaboration success in online communities: Explaining rises and falls from grace in Wikipedia. *MIS Q.* **35**, 613–628 (2011).
13. Bloom, N., *et al.* The impact of COVID-19 on productivity. Working Paper 28233, National Bureau of Economic Research, December 2020.
14. Chetty, R., Friedman, J. N., Hendren, N., Stepner, M., & Team, T. O. I. The economic impacts of COVID-19: Evidence from a new public database built using private sector data. Working Paper 27431, National Bureau of Economic Research, June 2020.
15. Suh, J., Horvitz, E., White, R. W. & Althoff, T. Population-scale study of human needs during the COVID-19 pandemic: Analysis and implications. arXiv:2008.07045 (arXiv preprint) (2020).
16. Desvars-Larrive, A. *et al.* A structured open dataset of government interventions in response to COVID-19. *medRxiv* **20**, 20 (2020).
17. Flaxman, S. *et al.* Estimating the effects of non-pharmaceutical interventions on COVID-19 in Europe. *Nature* **584**(7820), 257–261 (2020).
18. Zheng, Q. *et al.* HIT-COVID, a global database tracking public health interventions to COVID-19. *Sci. Data* **7**(1), 1–8 (2020).
19. Sultana, A., Tasnim, S., Bhattacharya, J., Hossain, M. M. & Purohit, N. Digital screen time during COVID-19 pandemic: A public health concern. osf.io/preprints/socarxiv/e8sg7 (2020).
20. Feldmann, A., *et al.* The lockdown effect: Implications of the COVID-19 pandemic on internet traffic. In *Proceedings of the ACM Internet Measurement Conference* (2020), pp. 1–18.

21. Al Tamime, R., Giordano, R., & Hall, W. Observing burstiness in Wikipedia articles during new disease outbreaks. In *Proceedings of the 10th ACM Conference on Web Science* (2018), pp. 117–126.
22. Zhang, A. F. *et al.* Participation of new editors after times of shock on Wikipedia. *Proc. Int. AAAI Conf. Web Soc. Media* **13**, 560–571 (2019).
23. Atchison, C. J. *et al.* Perceptions and behavioural responses of the general public during the COVID-19 pandemic: A cross-sectional survey of UK adults. *medRxiv* **20**, 20 (2020).
24. Jay, J. *et al.* Neighbourhood income and physical distancing during the COVID-19 pandemic in the United States. *Nat. Human Behav.* **2**, 1–9 (2020).
25. Suh, B., Convertino, G., Chi, E. H. & Pirolli, P. The singularity is not near: Slowing growth of Wikipedia. In *Proceedings of the 5th International Symposium on Wikis and Open Collaboration* (New York, NY, USA, 2009), WikiSym '09, Association for Computing Machinery.
26. Halfaker, A., Kittur, A., & Riedl, J. Don't bite the newbies: How reverts affect the quantity and quality of Wikipedia work. In *Proceedings of the 7th international symposium on wikis and open collaboration* (2011), pp. 163–172.
27. Xu, B. & Li, D. An empirical study of the motivations for content contribution and community participation in Wikipedia. *Inform. Manag.* **52**(3), 275–286 (2015).
28. Shaw, A. & Hargittai, E. The pipeline of online participation inequalities: The case of Wikipedia editing. *J. Commun.* **68**(1), 143–168 (2018).
29. Yasseri, T., Sumi, R. & Kertész, J. Circadian patterns of Wikipedia editorial activity: A demographic analysis. *PLoS One* **7**(1), e30091 (2012).
30. Robert, L. P. Jr. & Romero, D. M. The influence of diversity and experience on the effects of crowd size. *J. Am. Soc. Inf. Sci.* **68**(2), 321–332 (2017).
31. Butler, B., Joyce, E. & Pike, J. Don't look now, but we've created a bureaucracy: The nature and roles of policies and rules in Wikipedia. In *Proceedings of the SIGCHI Conference on human Hactors in Computing Systems* (2008), pp. 1101–1110.
32. Keegan, B. & Fiesler, C. The evolution and consequences of peer producing Wikipedia's rules. *Proc. Int. AAAI Conf. Web Soc. Media* **11**, 1 (2017).
33. Jemielniak, D. & Wilamowski, M. Cultural diversity of quality of information on wikipedias. *J. Assoc. Inf. Sci. Technol.* **68**(10), 2460–2470 (2017).
34. Miquel-Ribé, M. & Laniado, D. Cultural identities in Wikipedias. In *Proceedings of the 7th 2016 International Conference on Social Media and Society* (2016), pp. 1–10.
35. Scheffer, M., Westley, F. & Brock, W. Slow response of societies to new problems: Causes and costs. *Ecosystems* **6**(5), 493–502 (2003).
36. Scheffer, M. & Westley, F. R. The evolutionary basis of rigidity: Locks in cells, minds, and society. *Ecol. Soc.* **12**, 2 (2007).
37. Folke, C. *et al.* Resilience thinking: Integrating resilience, adaptability and transformability. *Ecol. Soc.* **15**, 4 (2010).
38. Zhang, A. F., Livneh, D., Budak, C., Robert, L. & Romero, D. Shocking the crowd: The effect of censorship shocks on Chinese Wikipedia. *Proc. Int. AAAI Conf. Web Soc. Media* **11**, 1 (2017).
39. Xiong, W. & Lagerström, R. Threat modeling—systematic literature review. *Comput. Secur.* **84**, 53–69 (2019).
40. Anthony, D., Smith, S. W. & Williamson, T. Reputation and reliability in collective goods: The case of the online encyclopedia wikipedia. *Ration. Soc.* **21**(3), 283–306 (2009).
41. Zheng, L., Albano, C. M., Vora, N. M., Mai, F. & Nickerson, J. V. The roles bots play in Wikipedia. *Proceedings of the ACM on Human–Computer Interaction* **3**, CSCW (2019), 1–20.
42. Yasseri, T., Sumi, R., Rung, A., Kornai, A. & Kertész, J. Dynamics of conflicts in Wikipedia. *PLoS One* **7**(6), e38869 (2012).
43. Sumi, R., Yasseri, T., et al. Edit wars in Wikipedia. In *2011 IEEE Third International Conference on Privacy, Security, Risk and Trust and 2011 IEEE Third International Conference on Social Computing* (2011), IEEE, pp. 724–727.
44. Kittur, A., Suh, B., Pendleton, B. A. & Chi, E. H. He says, she says: Conflict and coordination in Wikipedia. In *Proceedings of the SIGCHI Conference on Human Factors in Computing Systems* (2007), pp. 453–462.
45. Borra, E. et al. *Societal Controversies in Wikipedia Articles* 193–196 (Association for Computing Machinery, 2015).
46. Brandes, U., et al. Network analysis of collaboration structure in Wikipedia. In *WWW '09* (New York, NY, USA, 2009), Association for Computing Machinery, pp. 731–740.
47. Garcia, D. & Rimé, B. Collective emotions and social resilience in the digital traces after a terrorist attack. *Psychol. Sci.* **30**(4), 617–628 (2019).
48. Ciampaglia, G. L., & Taraborelli, D. MoodBar: Increasing new user retention in Wikipedia through lightweight socialization. In *Proceedings of the 18th ACM Conference on Computer Supported Cooperative Work and Social Computing* (2015), pp. 734–742.
49. Keegan, B. C., & Tan, C. A quantitative portrait of Wikipedia's high-tempo collaborations during the 2020 coronavirus pandemic. arXiv:2006.08899 (arXiv preprint) (2020).
50. Hill, B. M. & Shaw, A. The Wikipedia gender gap revisited: Characterizing survey response bias with propensity score estimation. *PLoS One* **8**(6), e65782 (2013).
51. Collier, B. & Bear, J. Conflict, criticism, or confidence: An empirical examination of the gender gap in Wikipedia contributions. In *Proceedings of the ACM 2012 Conference on Computer Supported Cooperative Work* (New York, NY, USA, 2012), CSCW '12, Association for Computing Machinery, pp. 383–392.
52. Chang, S. *et al.* Mobility network models of COVID-19 explain inequities and inform reopening. *Nature* **2**, 1–6 (2020).
53. Tsvetkova, M., García-Gavilanes, R., Floridi, L. & Yasseri, T. Even good bots fight: The case of Wikipedia. *PLoS One* **12**(2), e0171774 (2017).
54. Wikimedia Foundation. Analytics datasets: Mediawiki history. https://dumps.wikimedia.org/other/mediawiki_history/readme.html. Accessed 13 Dec 2020 (2020).
55. Wikimedia Statistics. Active editors by country with 5 to 99 edits (2021).
56. Diego Sáez-Trumper. COVID-19 Wikipedia data. https://covid-data.wmflabs.org. Accessed 13 Dec 2020 (2020).
57. Vrandečić, D. & Krötzsch, M. Wikidata: A free collaborative knowledgebase. *Commun. ACM* **57**(10), 78–85 (2014).
58. Wikimedia Foundation. REST API Documentation. https://wikimedia.org/api/rest_v1/. Accessed 1 Jan 2021 (2021).
59. Truong, C., Oudre, L. & Vayatis, N. Selective review of offline change point detection methods. *Signal Process.* **167**, 107299 (2020).
60. Gruber, J. The incidence of mandated maternity benefits. *Am. Econ. Rev.* **2**, 622–641 (1994).
61. Mian, A. & Sufi, A. House prices, home equity-based borrowing, and the US household leverage crisis. *Am. Econ. Rev.* **101**(5), 2132–56 (2011).
62. Gelman, A. & Huang, Z. Estimating incumbency advantage and its variation, as an example of a before-after study. *J. Am. Stat. Assoc.* **103**(482), 437–446 (2008).
63. Leys, C., Ley, C., Klein, O., Bernard, P. & Licata, L. Detecting outliers: Do not use standard deviation around the mean, use absolute deviation around the median. *J. Exp. Soc. Psychol.* **49**(4), 764–766 (2013).
64. Bilinski, A., & Hatfield, L. A. Nothing to see here? Non-inferiority approaches to parallel trends and other model assumptions (2020).
65. Olden, A. & Møen, J. The triple difference estimator. *NHH Dept. of Business and Management Science Discussion Paper*, 2020/1 (2020).

## Acknowledgements

Supported by TU Graz Open Access Publishing Fund. R.W.'s lab was partly funded by the Swiss National Science Foundation (grant 200021_185043), the European Union (TAILOR, grant 952215), Collaborative Research on Science and Society, and by generous gifts from Microsoft, Facebook, and Google.

## Author contributions

T.R. retrieved the dataset, processed the data, and performed the experiments. T.R. and M.H.R. wrote the code. T.R., M.H.R., and T.S. analyzed the data. T.R., M.H.R., T.S., F.L., M.S., R.W., and D.H. conceived and designed the experiments, developed the arguments, and wrote the paper.

## Competing interests

The authors declare no competing interests.

## Additional information

**Supplementary Information** The online version contains supplementary material available at https://doi.org/10.1038/s41598-021-00789-3.

**Correspondence** and requests for materials should be addressed to T.R.

**Reprints and permissions information** is available at www.nature.com/reprints.

**Publisher's note** Springer Nature remains neutral with regard to jurisdictional claims in published maps and institutional affiliations.